\documentclass{article}

\usepackage{arxiv}

\usepackage[utf8]{inputenc} 
\usepackage[T1]{fontenc}    
\usepackage{hyperref}       
\usepackage{url}            
\usepackage{booktabs}       
\usepackage{amsfonts}       
\usepackage{nicefrac}       
\usepackage{microtype}      
\usepackage{graphicx}
\usepackage{doi}
\usepackage{algorithmic}
\usepackage{bm}
\usepackage{amsmath}
\usepackage{doi}

\title{A Non-Negative Least Squares-based Approach for Moment-Preserving Particle Merging}

\author{\href{https://orcid.org/0000-0002-1434-7166}{\includegraphics[scale=0.06]{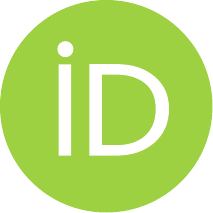}\hspace{1mm}Georgii Oblapenko$^1$}\thanks{
Preprint submitted to Proceedings of the 33rd International Symposium on Rarefied Gas Dynamics}\\
	$^1$Chair of Applied and Computational Mathematics, RWTH Aachen,
	Schinkelstrasse 52, 52062 Aachen, Germany \\
	$^\ast$Corresponding author. E-mail: \texttt{oblapenko@acom.rwth-aachen.de}
}

\renewcommand{\shorttitle}{A Non-Negative Least Squares-based Approach for Moment-Preserving Particle Merging}

\hypersetup{
pdftitle={A Non-Negative Least Squares-based Approach for Moment-Preserving Particle Merging},
pdfsubject={q-bio.NC, q-bio.QM},
pdfauthor={Georgii Oblapenko},
pdfkeywords={ Boltzmann equation, Direct Simulation Monte Carlo, particle merging, rarefied gas},
}

\begin{document}
\maketitle

\begin{abstract}
	In the present work, a novel particle merging scheme is proposed for PIC-DSMC simulations, based on the solution of a Non-negative Least Squares problem. The merging algorithm conserves arbitrary moments of the velocity distribution function, and a collision rate-conserving version of the algorithm is presented as well. Numerical simulations show excellent performance of the merging algorithm in terms of accuracy.
\end{abstract}


\section{Introduction}
Rarefied gas flows arise in a variety of engineering applications, such as aerospace engineering, semiconductor manufacturing, and nanotechnology applications~\cite{shen2006rarefied}. Their non-equilibrium nature precludes the use of traditional continuum-based computational fluid dynamics approaches for their simulation and necessitates the use of either higher-order moment equations~\cite{Torr}, or of kinetic solvers, such as the Direct Simulation Monte Carlo (DSMC) method~\cite{bird_1994}. The DSMC method has established itself as one of the main tools for rarefied gas dynamics simulations. In its standard formulation, the DSMC approach assumes that each computational particle represents a fixed number of actual particles. This causes issues in simulating flows with large density gradients and/or trace chemical species, as resolving the low populations  requires a large number of simulation particles.

To resolve this issue, variable-weight DSMC approaches have been suggested, where the weight of the computational particles are not fixed~\cite{rjasanow1996stochastic,boyd1996conservative,araki2020interspecies}. The drawback is that, during collisions, particles have to be split, leading to an exponential growth in the number of particles if left unchecked. The growth becomes even more rapid if modern collision schemes with improved resolution of low-probability events are used~\cite{araki2020interspecies,oblapenko2022hedging}.
Therefore, particle merging is required, i.e. a procedure that reduces the number of particles in a simulation. This inherently incurs an information loss and may lead to significant numerical errors. Multiple merging algorithms have been developed~\cite{rjasanow1998reduction,vranic2015particle,martin2016octree,chen2024stochastic} in an attempt to reduce this error and improve the efficiency of the merging procedure. They frequently rely on a ``divide-and-conquer`` strategy: the particles chosen for merging are grouped into smaller sets in velocity space, and in each of these subsets all the particles are replaced by a very small number (usually 1 or 2) of particles.
The advantage of such an approach is that it is possible to compute the weights, positions, and velocities of the post-merge particles in each subset analytically. However, the analytical merging of particles is restricted to conservation of lower-order moments of particle distribution, as conservation of higher-order moments requires solving of systems of non-linear equations. In addition, it is in general not possible to guarantee that post-merge particles will not end up outside of the physical domain or that their velocities will not exceed the velocities of the original set of particles. Therefore, it is of interest to design merging schemes that can ensure conservation of higher-order moments and constraint satisfaction~\cite{lama2020higher}.
The recent scheme of Gonoskov~\cite{gonoskov2022agnostic} is an example of a moment-preserving merging scheme that has also been successfully coupled with the octree grouping approach~\cite{huerta2024situ}. In this work we investigate an alternative moment-preserving merging scheme based on solution of a non-negative least squares problem and apply it to simulation of two model spatially homogeneous problems. In addition, a rate-preserving version of the scheme is formulated for plasma simulations. We compare the results with those obtained by the octree merging scheme and show that the new scheme leads to lower bias in the numerical solution.

\section{Particle Merging}
We consider a set of particles $\mathcal{P}_N = \{(w^{(i)},v^{(i)}_{x},v^{(i)}_{y},v_z^{(i)})\}_{i=1}^N$, where each particle is characterized by its non-negative computational weight $w^{(i)} \geq 0$ and the 3 velocity components $v^{(i)}_{x}$, $v^{(i)}_{y}$, $v_z^{(i)}$. In the present work we disregard the particle positions and focus only on preservation of velocity moments.

Merging of particles corresponds to computation of a new set of particles $\hat{\mathcal{P}}_M = \{(\hat{w}^{(i)},\hat{v}^{(i)}_{x},\hat{v}^{(i)}_{y},\hat{v}_z^{(i)})\}_{i=1}^M$, $M<N$.

We are interested in conserving central moments, but since we assume Galilean invariance, once can always compute the mean velocities $\overline{v}_x$, $\overline{v}_y$, $\overline{v}_z$, subtract them from the velocities of $\mathcal{P}_N$, merge down the particles, and add the mean velocity components to the post-merge particles $\hat{\mathcal{P}}_M$. Therefore, we assume that $\overline{v}_x=\overline{v}_y=\overline{v}_z=0$ and focus on the non-central moments
\begin{equation}
    M_{jkl}(\mathcal{P}) = \frac{1}{\sum_w} \sum_i  w^{(i)} \left(v^{(i)}_{x} \right)^j \left(v^{(i)}_{y} \right)^k \left(v^{(i)}_{z}\right)^l.\label{eq:moments}
\end{equation}

We are interested in developing a merging approach that can conserve not only the basic invariants of mass, momentum, and energy, but also certain higher-order moments $M_{jkl}$ of prescribed order $j,k,l$.

\subsection{Non-Negative Least Squares Merging}
Now we consider an alternative merging algorithm, based on the non-negative least squares approach. Let us consider the definition~(\ref{eq:moments}) for a set of moments $(M_{j_1k_1l_1},\ldots,M_{j_Lk_Ll_L})$ in matrix form:
\begin{equation}
    \mathbf{V}\mathbf{w} = \mathbf{M},\label{eq:moment-system-matrix}
\end{equation}
where the matrices and vectors $\mathbf{V} \in \mathbb{R}^{L \times N}$, $\mathbf{w} \in \mathbb{R}^{N}$, $\mathbf{M} \in \mathbb{R}^{L}$ are defined as
\begin{equation}
\mathbf{V} = 
\begin{pmatrix}
\left(v^{(1)}_{x} \right)^{j_1} \left(v^{(1)}_{y} \right)^{k_1} \left(v^{(1)}_{z}\right)^{l_1} & \ldots & \left(v^{(N)}_{x} \right)^{j_1} \left(v^{(N)}_{y} \right)^{k_1} \left(v^{(N)}_{z}\right)^{l_1}\\
 & \ddots &\\
\left(v^{(1)}_{x} \right)^{j_L} \left(v^{(1)}_{y} \right)^{k_L} \left(v^{(1)}_{z}\right)^{l_L} & \ldots & \left(v^{(N)}_{x} \right)^{j_L} \left(v^{(N)}_{y} \right)^{k_L} \left(v^{(N)}_{z}\right)^{l_L}
\end{pmatrix},\label{eq:def-m-v}
\end{equation}
\begin{equation}
\mathbf{w} =
    \begin{pmatrix}
     w^{(1)} \ldots
     w^{(N)}
\end{pmatrix}^{\mathrm{T}},\quad
  \mathbf{M}   =
\begin{pmatrix}
     M_{j_1k_1l_1} \ldots
     M_{j_Lk_Ll_L}
\end{pmatrix}^{\mathrm{T}}.\label{eq:def-m-M}
\end{equation}
We can switch our point of view and consider~(\ref{eq:moment-system-matrix}) not as a definition (i.e. given particle velocities and weights, one can compute the moments), but as a system of linear equations for $\mathbf{w}$, i.e. the particle weights: given a set of moments and particle velocities, one can compute the weights. Depending on the number of particle velocities and moments, this system is be either under- ($N > L$), well- ($N=L$, or over-determined ($N<L$).
We consider the under-determined case ($N>L$), that is, we have more unknown particle weights than moments. However, we augment the system with the requirement that $w_i \geq 0$. This leads directly to a non-negative least squares (NNLS) problem~\cite{lawson1995solving}. One property of solutions of under-determined NNLS problems is the sparsity of the solution: that is, it is highly likely that some entries of the solution vector $\mathbf{w}$ will be 0. We can leverage this sparsity property of NNLS solutions in order to use the NNLS algorithm as a merging method:
\begin{enumerate}
    \item Given a set of particles $\mathcal{P}_N$, we compute the matrix $\mathbf{V}$ and vector $\mathbf{M}$, as defined by~(\ref{eq:def-m-v})--(\ref{eq:def-m-M})
    \item Next, we solve system~(\ref{eq:moment-system-matrix}) for $\mathbf{w}'$ using the NNLS method
    \item We replace the original set of particles $\mathcal{P}_N$ with new particles, by taking the velocities from the original set of velocities used to construct $\mathbf{V}$ and new weights from the solution vector $\mathbf{w}$, skipping creation of particles whose weights are equal to 0.
\end{enumerate}
Thus one can can conserve any velocity and spatial moments (not considered in the present work, but incorporating them only means adding additional rows to $\mathbf{V}$ and $\mathbf{M}$), and also has full control over the post-merge velocities and spatial locations of the particles, since they are explicitly used to construct $\mathbf{V}$.
However, the existence of a solution is not guaranteed. In addition, we have no control of the number of non-zero weights, although in practice it is usually equal to $L$.

To increase the chances of finding a solution, one can add additional columns to the matrix $\mathbf{V}$, by adding points in velocity space and computing their contributions to the moments. One can choose these points so as to avoid their velocities exceeding the velocity bounds of the original set of particles $\mathcal{P}_N$, for example by only adding points from inside the convex hull of $\mathcal{P}_N$. But this procedure is computationally expensive and therefore, a faster approach is used in the present work which adds less rigorous bounds on these additional velocities.

We compute the velocities defined by the second-order moments: $v_{x,M_2}=\sqrt{M_{200}}$, $v_{y,M_2}=\sqrt{M_{020}}$, $v_{z,M_2}=\sqrt{M_{020}}$. We then augment the matrix $\mathbf{V}$ with additional columns computed using new velocities with a magnitude of $\alpha v_{i,M_2}$, $i =x,y,z$, where $\alpha \in [0,1]$ is a user-defined multiplier, with the velocity signs chosen so that these additional velocities cover all octants in velocity space.
In the present work, 16 additional points columns are added to $\mathbf{V}$ in this manner: 8 points with $\alpha=1$ and 8 points with $\alpha=0.5$. An additional constraint is added that the velocities of these points do not exceed the minimum and maximum velocities in the corresponding directions. This has been found to improve the stability of the algorithm, but a more detailed investigation of the optimal choice of additional velocities is left for future work. If on a given timestep the NNLS merging approach fails to produce a sparse solution to the linear system~(\ref{eq:moment-system-matrix}), one can switch to an alternative merging approach.

We also consider conservation of inter-species collision rates. This is complicated by the fact that collision rates depend on the relative velocity of the colliding particles $g$, i.e. on both species' VDFs. However, for the specific case of electron-neutral collisions in a plasma, one can use the common assumption that the electron velocities are significantly higher than those of the neutrals:
\begin{equation}
    v_{\mathfrak{e}} \gg v_{\mathfrak{n}},
\end{equation}
where $ v_{\mathfrak{e}}$ is the velocity of electrons, and $v_{\mathfrak{n}}$ is the velocity of the neutrals. Using this assumption, one can write for an electron-neutral collision rate $k$ between a set of neutral and electron particles
\begin{equation}
    k(g) = \frac{1}{w_{\mathfrak{e}}w_{\mathfrak{n}}} \sum_i \sum_j w_{\mathfrak{e}}^{(i)}w_{\mathfrak{n}}^{(j)} |v_{\mathfrak{e}}^{(i)}-v_{\mathfrak{n}}^{(j)}| \sigma(|v_{\mathfrak{e}}^{(i)}-v_{\mathfrak{n}}^{(j)}|) \approx  \frac{1}{w_{\mathfrak{e}}}\sum_i w_{\mathfrak{e}}^{(i)} |v_{\mathfrak{e}}^{(i)}| \sigma(|v_{\mathfrak{e}}^{(i)}|).\label{eq:approx-k}
\end{equation}
Here $\sigma(g)$ is the cross-section of the process for which the rate is being calculated and is a function of the relative velocity of the colliding particles.
With this assumption, the approximate rate becomes a function of the electron VDF only, and can thus be conserved when merging the electron particles by adding the following row to matrix $\mathbf{V}$:
\begin{equation}
   \left(|v_{\mathfrak{e}}^{(i)}| \sigma(|v_{\mathfrak{e}}^{(i)}|) \ldots |v_{\mathfrak{e}}^{(N)}| \sigma(|v_{\mathfrak{e}}^{(N)}|) \right)\label{eq:kapprox_cons1}
\end{equation}
and the corresponding entry to the right-hand side vector $\mathbf{M}$:
\begin{equation}
    \sum_i w_{\mathfrak{e}}^{(i)} |v_{\mathfrak{e}}^{(i)}| \sigma(|v_{\mathfrak{e}}^{(i)}|).\label{eq:kapprox_cons2}
\end{equation}
This procedure can be carried out for multiple different processes, such as elastic scattering, electron-impact electronic excitation, electron-impact ionization, etc.

\section{Numerical Results}
The octree merging and NNLS-based merging approaches were implemented in Merzbild.jl, a DSMC code implemented in the Julia programming language. The code and the input files used to produce the results for the present work are publicly available on Github~\cite{oblapenko2024merzbild}. As a baseline merging approach, the octree binning algorithm~\cite{martin2016octree} was used.

\subsection{BKW Relaxation}
We first consider the BKW relaxation problem~\cite{bobylev1976,krook1977}, as we can compare the numerical solutions to the analytical solution.

The problem is initialized by evaluating the velocity distribution function on a fine 32$^3$ velocity grid and converting the the points on the grid with their associated VDF values to variable-weight particles. For NNLS merging, different numbers of conserved central moments were considered, with \textbf{all} mixed moments of order up to $L_{m,1}$ conserved (that is, all moments $M_{j_ik_il_i}$ such that $j_i + k_i + l_i \leq L_{m,1}$), and additional conservation of moments $M_{j_i00}$, $M_{0j_i0}$, $M_{0j_i0}$, where $L_{m,1} < j_i \leq L_{m,2}$. The target number of particles for the octree merging algorithm was chosen to be similar to that of the post-merge particles obtained via the NNLS merging. 400 simulations were performed for 500 timesteps each for each set of merging settings and then averaged, to reduce the impact of stochastic noise on the analysis of the results.

As an error metric, we consider the bias of the ensemble-averaged solution w.r.t. the analytical solution:
\begin{equation}
    \mathcal{B}\left(\hat{M}^{2l}\right) =\sqrt{ \frac{1}{N_t}  \sum_{t_i} \left(\overline{\hat{M}^{2l}}(t_i) - \hat{M}^{2l}_{an}(t_i)\right)^2 }.
\end{equation}
Here $\hat{M}^{2l}$ is the total moment of order $2l$ (the moment of the particle \textbf{speed} of order $2l$) for which we are considering the bias, $N_t$ is the number of timesteps for which the bias is computed, $t_i = i\Delta t$ is the time at timestep $i$, $\overline{\hat{M}^{2l}}$ is the ensemble average of the moments computed with the given set of simulation parameters, and $\hat{M}^{2l}_{an}$ is the analytical value of the moment.

\begin{figure}[h]
  \centering
  \includegraphics[width=0.475\textwidth]{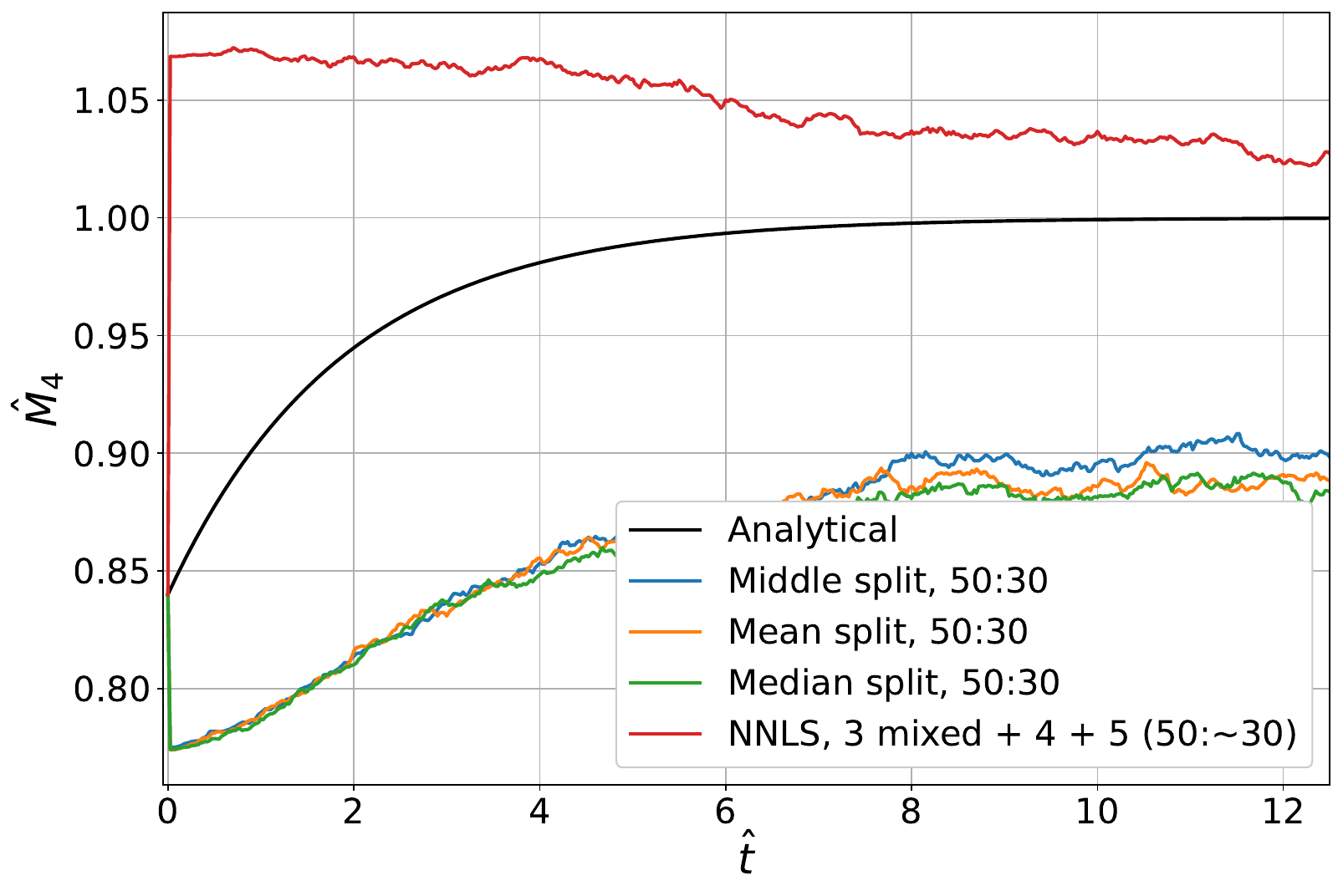}
  \includegraphics[width=0.475\textwidth]{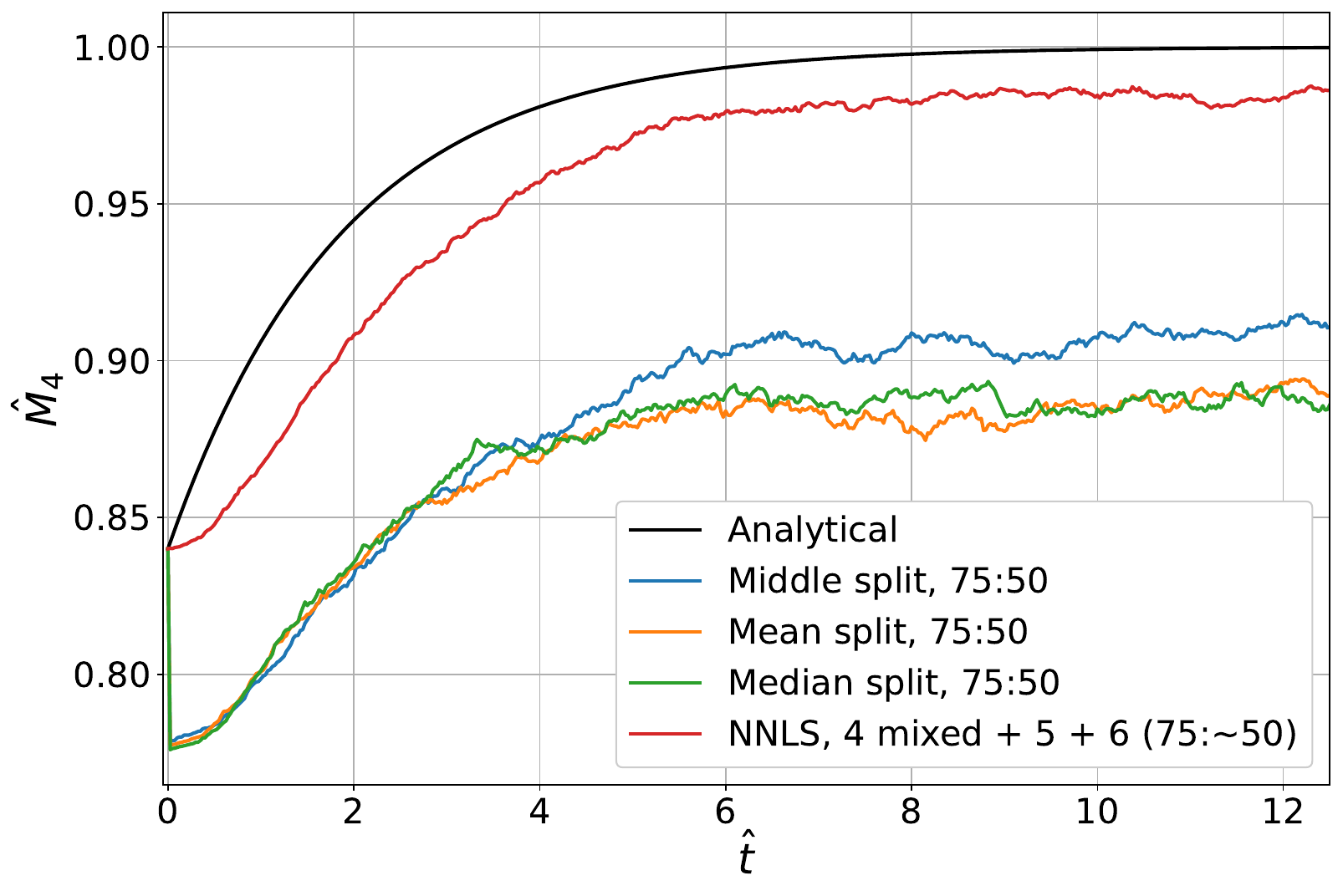}
  \caption{Evolution of the 4$^{th}$ total moment of the BKW distribution, 50:30 merge (left) and 75:50 merge (right).}
  \label{fig:m4_evolution}
\end{figure}

Figure~\ref{fig:m4_evolution} shows the ensemble-averaged evolution of the 4$^{th}$ total moment of the BKW distribution for different merging algorithms and different numbers of particles consider. The $N$:$M$ numbers on the legend denote the threshold number of particles $N$ (if it is exceeded, particle merging is performed) and target post-merge number of particles $M$. The ``middle split'', ``mean split'', and ``median split'' are different versions of the octree merging algorithm with different splitting of the bins for refinement: either along the middle velocity (regardless of the particles' velocities), along the mean velocity of the bin, or along the median velocity of the bin. On the left subplot, we see significant deviations from the analytical solution, and both the octree merging and NNLS-based merging algorithms exhibit a large jump in the moment at $t=0$, as the very finely sampled distribution is merged down to only a few dozen particles. On the right subplot, the octree merging algorithm exhibits similar behaviour, although the error is somewhat lower, as the number of particles is increased. The NNLS merging algorithm however now conserves a sufficient number of moments of the VDF to be able to preserve the 4$^{th}$ total moment during merging --- no jump can be seen at $t=0$, and the solution in general is much closer to the analytical one.

\begin{figure}[h]
  \centering
  \includegraphics[width=0.475\textwidth]{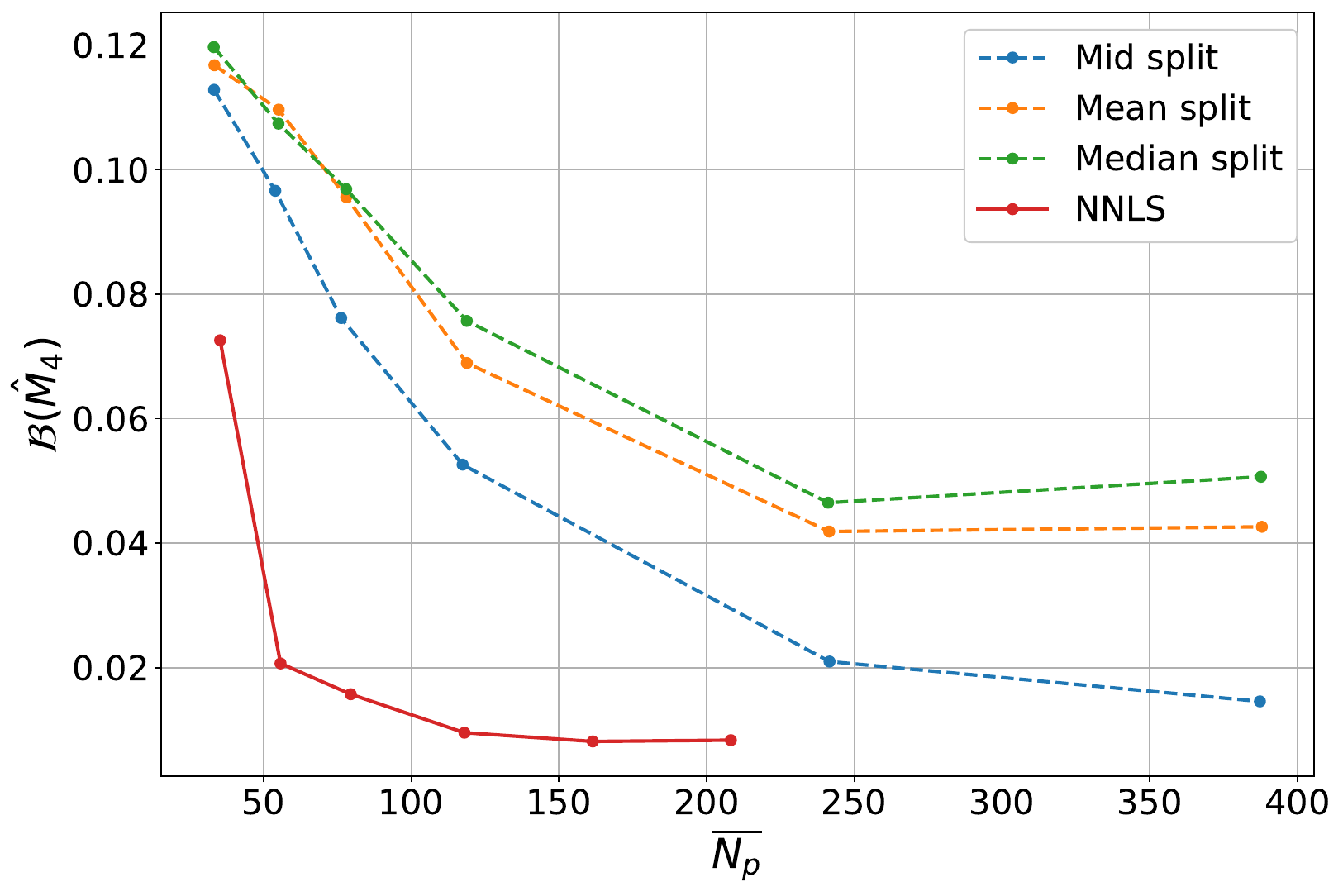}
  \caption{Bias in the 4$^{th}$ total moment of the BKW distribution as a function of the average number of particles.}
  \label{fig:m4_bias}
\end{figure}

Figure~\ref{fig:m4_bias} shows the bias in the 4$^{th}$ total moment of the BKW distribution as a function of the time-averaged number of particles $\overline{N}_p$. We see that all versions of the octree merging exhibit a significantly higher bias than the proposed moment-preserving approach, requiring at least 5 times as many particles to achieve a similar level of solution bias. Out of all the octree bin splitting strategies, the ``middle split'' performs the best, probably due to the split not being influenced by the distribution of particles in phase space and thus avoiding any induced bias.

\subsection{0-D Ionization}
Next, we consider a spatially homogeneous ionization in an argon plasma driven by a constant electric field of 400~Tn. Only electron-neutral elastic scattering and electron-impact ionization were accounted for, with isotropic scattering, and the collision cross-sections taken from the IST-Lisbon database~\cite{alves2014lisbon}.
After an initial burn-in time, the system reaches a quasi-steady state, characterized by a constant ionization rate coefficient. The event-splitting scheme was used for collisions~\cite{oblapenko2022hedging}. We analyze the bias introduced into the ionization rate coefficient in the first 100 timesteps after merging of the electron VDF is performed (averaged over all merging events). Since the system is driven by an external force, it has time to ``recover'' between merging events, and thus the bias immediately after merging has been performed is a better metric to compare the error introduced by the different merging approaches.

\begin{figure}[h]
  \centering
  \includegraphics[width=0.475\textwidth]{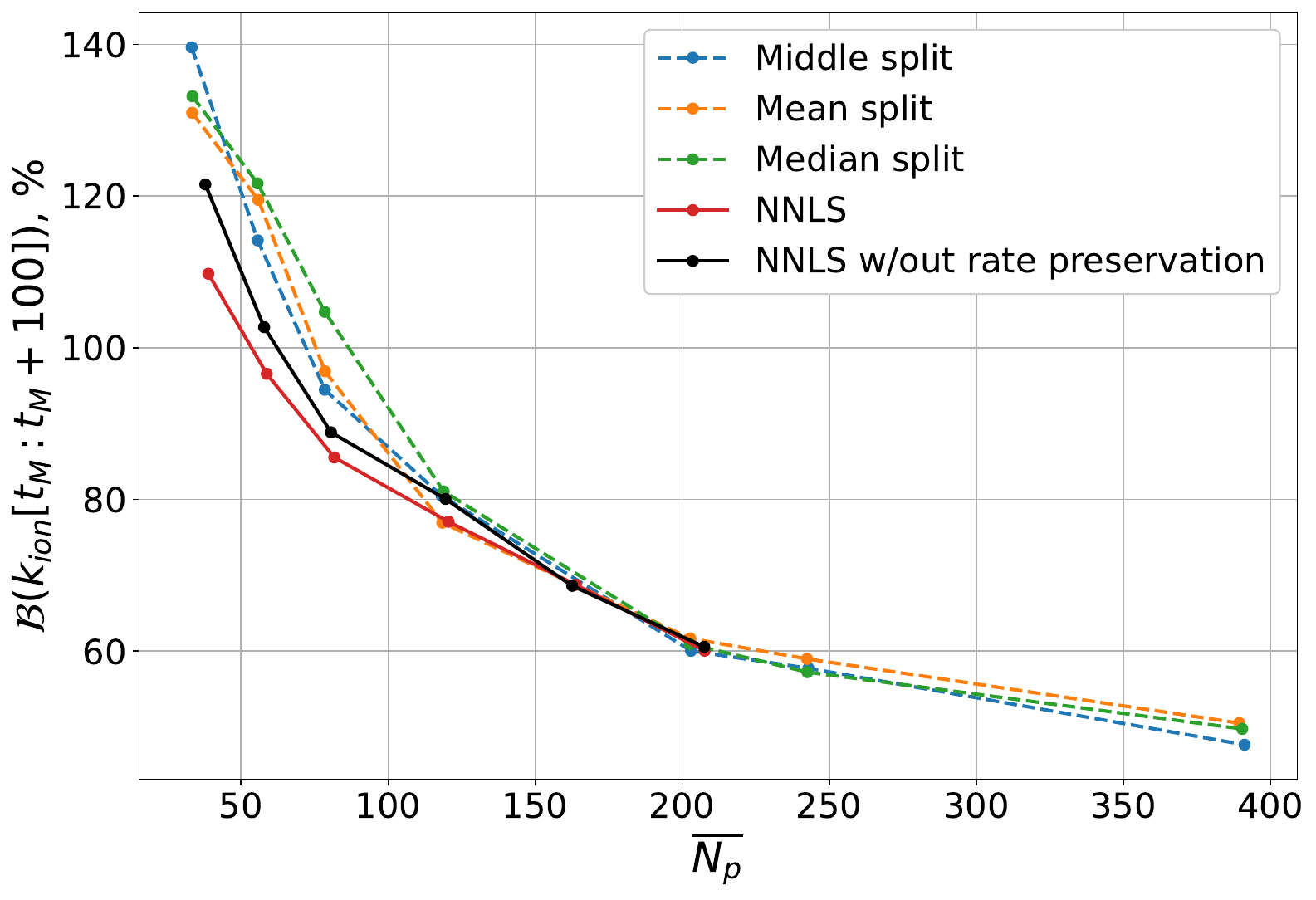}
  \caption{Average Bias in the ionization rate coefficient in the first 100 timesteps after merging events as a function of the average number of particles.}
  \label{fig:kion_bias}
\end{figure}
Figure~\ref{fig:kion_bias} shows the average bias in the ionization rate coefficient immediately after merging events. Two versions of the NNLS merging approach were considered: with and without preservation of approximate electron-neutral collision rates (red and black lines, correspondingly).
We see that for low numbers of particles, the NNLS-based approaches outperform the octree merging approach. The addition of the approximate rate preservation constraints also reduces the bias; however, as more and more moments are conserved in the NNLS merging algorithm (corresponding to a larger average number of particles $\overline{N_p}$), the difference between the rate preserving and non-rate preserving formulations becomes negligible, as the high-velocity tails of the electron VDF are sufficiently well-preserved during merging when $\overline{N_p}$ is sufficiently large.

\section{Conclusions}
A novel moment-preserving particle merging approach based on the solution of a non-negative least-squares problem has been developed, along with an approximate rate-conserving version for plasmas. It has been shown to noticeably reduce the merging-induced the bias in higher-order moments of the velocity distribution function, as well as provide a better representation of the ionization rate coefficient for low numbers of computational particles.

Future planned extensions to the work include coupling the NNLS approach with the octree merging algorithm~\cite{martin2016octree} for a greater degree of adaptivity, and consideration of spatial moments for non-spatially-homogeneous problems.

\section*{Acknowledgments}
This work has been supported by the German Research Foundation within the research unit SFB 1481.


\end{document}